\documentclass[twocolumn,floatfix,superscriptaddress,prb]{revtex4}
\usepackage{graphicx}
\usepackage{amsmath}
\usepackage{amssymb}
\usepackage{bm}

\begin{document}

\title{Disorder and magnetic-field induced breakdown of helical edge conduction in an inverted electron-hole bilayer}
\author{D. I. Pikulin}
\affiliation{Instituut-Lorentz, Universiteit Leiden, P.O. Box 9506, 2300 RA Leiden, The Netherlands}
\author{T. Hyart}
\affiliation{Instituut-Lorentz, Universiteit Leiden, P.O. Box 9506, 2300 RA Leiden, The Netherlands}
\author{Shuo Mi}
\affiliation{Instituut-Lorentz, Universiteit Leiden, P.O. Box 9506, 2300 RA Leiden, The Netherlands}
\author{J. Tworzyd{\l}o}
\affiliation{Institute of Theoretical Physics, Faculty of Physics, University of Warsaw, Ho\.{z}a 69, 00--681 Warsaw, Poland}
\author{M. Wimmer}
\affiliation{Kavli Institute of Nanoscience, Delft University of Technology, P.O. Box 5046, 2600 GA Delft, The Netherlands}
\author{C. W. J. Beenakker}
\affiliation{Instituut-Lorentz, Universiteit Leiden, P.O. Box 9506, 2300 RA Leiden, The Netherlands}
\date{April 2014}
\begin{abstract}
We calculate the conductance of a two-dimensional bilayer with inverted electron-hole bands, to study the sensitivity of the quantum spin Hall insulator (with helical edge conduction) to the combination of electrostatic disorder and a perpendicular magnetic field. The characteristic breakdown field for helical edge conduction splits into two fields with increasing disorder, a field $B_{c}$ for the transition into a quantum Hall insulator (supporting chiral edge conduction) and a smaller field $B'_{c}$ for the transition to bulk conduction in a quasi-metallic regime. The spatial separation of the inverted bands, typical for broken-gap InAs/GaSb quantum wells, is essential for the magnetic-field induced bulk conduction --- there is no such regime in HgTe quantum wells.
\end{abstract}
\maketitle

A two-dimensional band insulator can support two types of conducting edge states: counterpropagating (helical) edge states in zero magnetic field and unidirectional (chiral) edge states in a sufficiently strong perpendicular field. These two topologically distinct phases are referred to as a quantum spin Hall (QSH) and quantum Hall (QH) insulator, respectively.\cite{Has10,Xi11} The physics of the QSH-to-QH transition is governed by band inversion:\cite{Tka10,Che12,Sch12} The electron-like and hole-like subbands near the Fermi level are interchanged in a QSH insulator, so that the band gap in the bulk becomes smaller rather than larger with increasing perpendicular magnetic field.\cite{Mey90,Sch98} The gap closing at a characteristic field $B_{c}$ signals the transition from an inverted QSH gap with helical edge states to a non-inverted QH gap supporting chiral edge states.

\begin{figure}[tb]
\centerline{\includegraphics[width=0.8\linewidth]{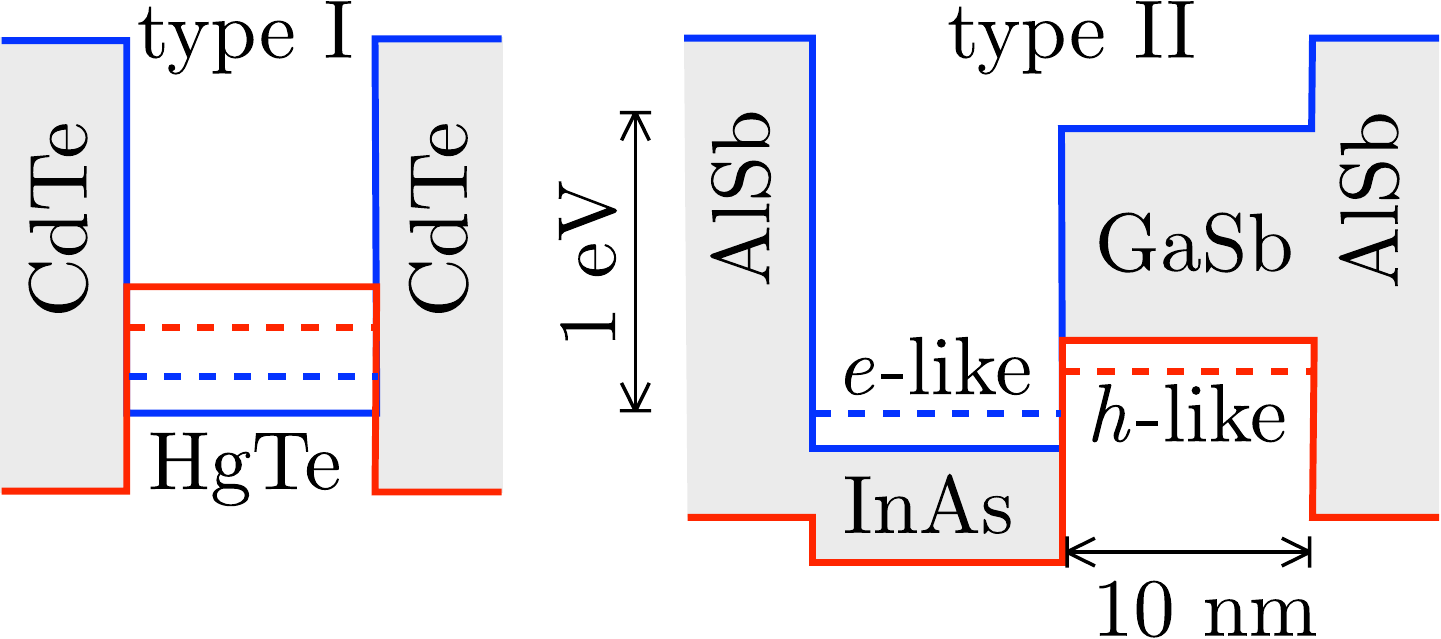}}
\caption{Alignment of conduction band (blue) and valence band (red) in a quantum well of type I (panel \textit{a}) and type II (panel \textit{b}). Both quantum wells have electron and hole subbands in inverted order (dotted lines, red \textit{h}-like above blue \textit{e}-like). The band gap (grey) is broken in the InAs/GaSb quantum well of type II, providing for a spatially separated electron-hole bilayer. There is no such spatial separation in the HgTe quantum well of type I.}
\label{fig_bandalign}
\end{figure}

The early experiments on the QSH effect were performed in HgTe layers with CdTe barriers (type I quantum wells).\cite{Ber06,Kon07} Recently the effect has also been observed in InAs/GaSb bilayers with AlSb barriers (type II quantum wells).\cite{Liu08a,Kne09,Suz13,Du13} Both types of quantum wells can have electron-hole subbands in inverted order, but while these are strongly coupled in type I quantum wells, they are spatially separated and weakly coupled in the broken-gap quantum wells of type II (see Fig.\ \ref{fig_bandalign}). Although the difference has no consequences in zero magnetic field, we will show here that the breakdown of helical edge conduction in a magnetic field becomes qualitatively different.

In both type I and type II quantum wells we find an increase with disorder of the characteristic field $B_{c}$ for the QSH-to-QH transition, as a consequence of the same mechanism that is operative in topological Anderson insulators:\cite{Li09} a disorder-induced renormalization of the band gap.\cite{Gro09} Basically, in a narrow-gap semiconductor the effect of disorder on the bulk band gap is opposite in the inverted and non-inverted case. While a non-inverted band gap is reduced by disorder, the inverted band gap is increased. Since $B_{c}$ is proportional to the zero-field band gap, it is pushed to larger fields by impurity scattering.

As a consequence, disorder increases the robustness of helical edge conduction in type I quantum wells, such as HgTe. In contrast, we find that in broken-gap quantum wells of type II a second transition at a weaker field $B'_{c}$ appears, at which helical edge conduction gives way to bulk conduction. This lower characteristic field splits off from $B_{c}$ with increasing disorder, producing a quasi-metallic regime in a broad field interval $B'_{c}\lesssim B\lesssim B_{c}$. The robustness of helical edge conduction is therefore reduced by disorder in type II quantum wells, such as InAs/GaSb. We discuss the magnetic-field induced bulk conduction in terms of Landau-level hybridization\cite{Suz04} and explain why it is only operative for weakly coupled electron-hole subbands.

Our investigation is based on the Bernevig-Hughes-Zhang Hamiltonian for inverted electron-hole bilayers.\cite{Ber06,Liu08a,Liu08b} In zero magnetic field the Hamiltonian of the clean system takes the form
\begin{align}	
&{\cal H}(\bm{k})=\begin{pmatrix}
H_{0}(\bm{k})&H_{1}(\bm{k})\\
-H_{1}^{\ast}(-\bm{k})&H_{0}^{\ast}(-\bm{k})
\end{pmatrix},\label{calHdef}\\
&H_{0}(\bm{k})=\begin{pmatrix}
M_{0}+\mu_{+}k^{2}&\beta k_{+}\\
\beta k_{-}&-M_{0}-\mu_{-}k^{2}
\end{pmatrix},\label{H0def}\\
&H_{1}(\bm{k})=\begin{pmatrix}
\Delta_{+}k_{+} -i\alpha k_{-}&-\Delta_0\\
\Delta_0&\Delta_{-}k_{-}
\end{pmatrix},\label{H1def}
\end{align}
as a function of wave vector $\bm{k}=(k_x,k_y)$ in the $x$-$y$ plane of the quantum well. We have defined $k^2=k_x^2+k_y^2$, $k_{\pm}=k_x\pm ik_y$, $\mu_{\pm}=\mu_{0}\pm\delta\mu$. It is a tight-binding Hamiltonian in the spin ($\uparrow\downarrow$) and subband ($\pm$) degrees of freedom, acting on a wave function with elements $(\psi_{+\uparrow},\psi_{-\uparrow},\psi_{+\downarrow},\psi_{-\downarrow})$. The term $\beta k_{\pm}$ in block $H_{0}$ describes the inter-subband coupling, and the block $H_{1}$ accounts for Rashba and Dresselhaus spin-orbit coupling. To model the two types of quantum wells we use the parameters listed in Table \ref{Tablequantumwell}.\cite{note1} 

\begin{table}
\begin{tabular}{c | c | c  }
  & type I & type II\\ \hline
$M_0 $ [eV]          & $-0.01$ &     $-0.01$\\
$\mu_{0}$ [eV$\cdot$\AA$^2$]     & $68.6$ &        $81.9$  \\
$\delta\mu$ [eV$\cdot$\AA$^2$]     & $51.1$ &   $21.6$ \\
$\beta$ [eV$\cdot$\AA]         & $3.65$ &      $0.72$  \\
$\Delta_{0}$  [eV]    & $0.0016$ & $0.0003$  \\
$\Delta_{+}$ [eV$\cdot$\AA] & $-0.128$ &     $0.0011$ \\
$\Delta_{-}$  [eV$\cdot$\AA]  & $0.211$ &    $0.0006$ \\
$\alpha$ [eV$\cdot$\AA]         & $0.0$ &   $0.16$ \\
\end{tabular}
\caption{Parameters of the tight-binding Hamiltonian \eqref{calHdef} used in the numerical simulations of quantum wells of type I (HgTe) and type II (InAs/GaSb).\cite{note1} The electron-hole asymmetry parameter $\delta\mu$ is set to zero in some of the calculations.
\label{Tablequantumwell}
}
\end{table}

The time-reversal symmetry breaking effect of a perpendicular magnetic field $\bm{B}=(0,0,B)$ is predominantly orbital, accounted for by the substitution $\bm{k}\mapsto \bm{k}-(e/\hbar)\bm{A}$, with vector potential $\bm{A}=(0,Bx,0)$. The Zeeman effect, which would be the dominant effect in a parallel field, is not included. (We will return to this later on.)

\begin{figure}[tb]
\centerline{\includegraphics[width=0.8\linewidth]{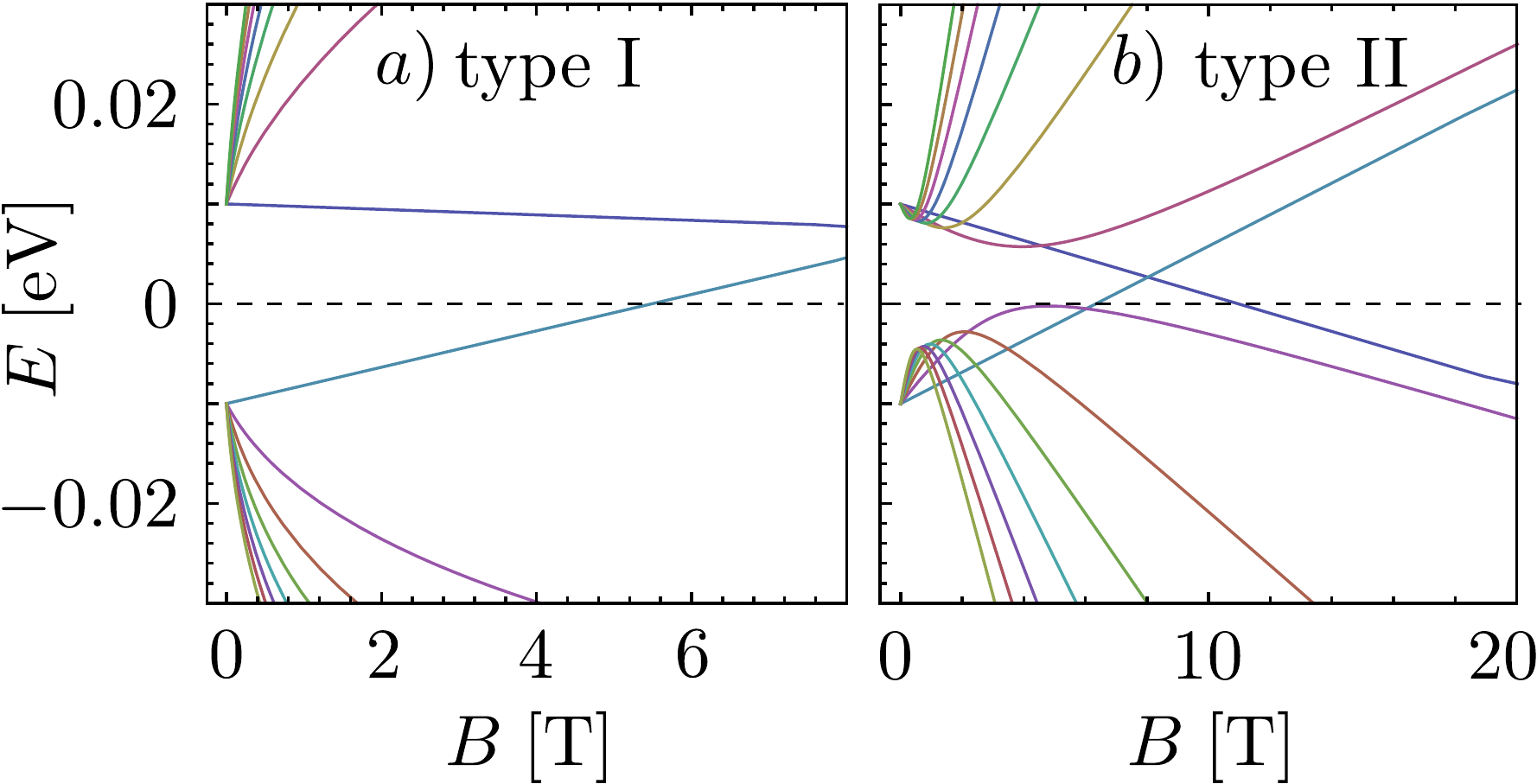}}
\caption{Landau level spectrum in the two types of quantum wells, calculated from the Hamiltonian \eqref{calHdef} for the parameters of Table \ref{Tablequantumwell} (nonzero $\delta\mu$ and $M_0=-0.01\,{\rm eV}$).}
\label{fig_LL1}
\end{figure}

\begin{figure}[tb]
\centerline{\includegraphics[width=0.6\linewidth]{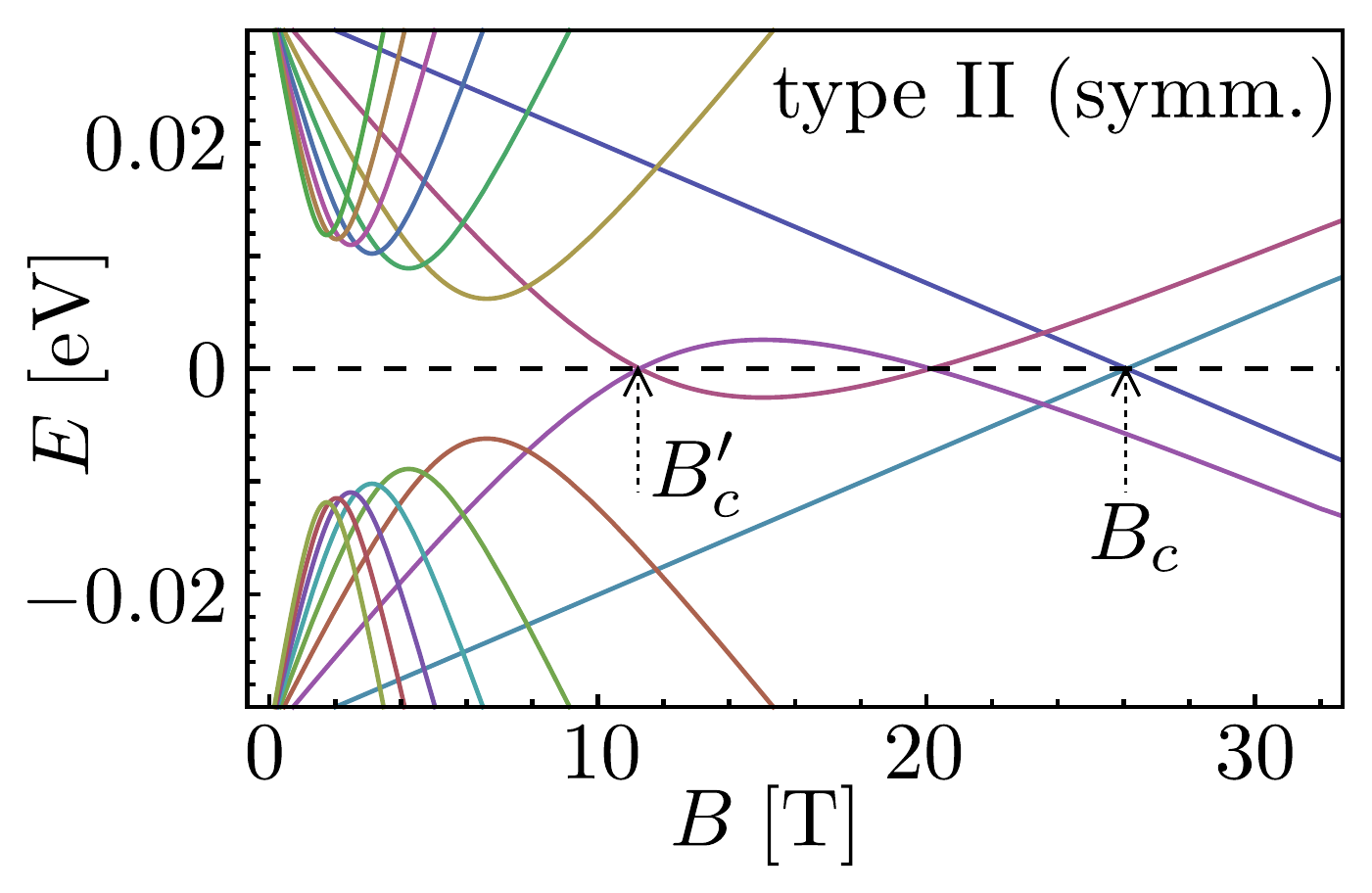}}
\caption{Same as Fig.\ \ref{fig_LL1}, for a type II quantum well with electron-hole symmetry ($\delta\mu=0$) and for a larger zero-field gap $M_0=-0.0325\,{\rm eV}$. The two characteristic fields $B'_c$ and $B_c$ of the first and last Landau level crossing are indicated.}
\label{fig_LL2}
\end{figure}

In Figs.\ \ref{fig_LL1} and \ref{fig_LL2} we show the magnetic field dependence of the Landau levels in the two types of quantum wells. If the inverted electron and hole subbands would be totally uncoupled, then all Landau levels from the valence band would move upwards while all Landau levels from the conduction band would move downwards --- resulting in an accumulation of Landau levels inside the zero-field band gap $|E|<|M_0|$. Electron-hole coupling hybridizes the Landau levels from conduction and valence band,\cite{Suz04} pushing them out of the gap. In a type I quantum well only a single pair of Landau levels remains inside the gap, see Fig.\ \ref{fig_LL1}\textit{a}. The spatial separation of the electron-hole subbands in a type II quantum well does allow for multiple Landau levels inside the gap, the more so the larger $|M_0|$ --- compare Figs.\ \ref{fig_LL1}\textit{b} and \ref{fig_LL2}.

To define the characteristic fields mentioned in the introduction it is convenient to set the electron-hole asymmetry parameter $\delta\mu$ to zero, so that the Landau level crossings are all in the middle of the gap, at $E=0$. As indicated in Fig.\ \ref{fig_LL2}, the first and the last level crossing then identify, respectively, $B'_c$ and $B_c$. As we will now show, these two fields delimit a regime of bulk conduction in a disordered type II quantum well.

\begin{figure*}[tb]
\centerline{\includegraphics[width=0.8\linewidth]{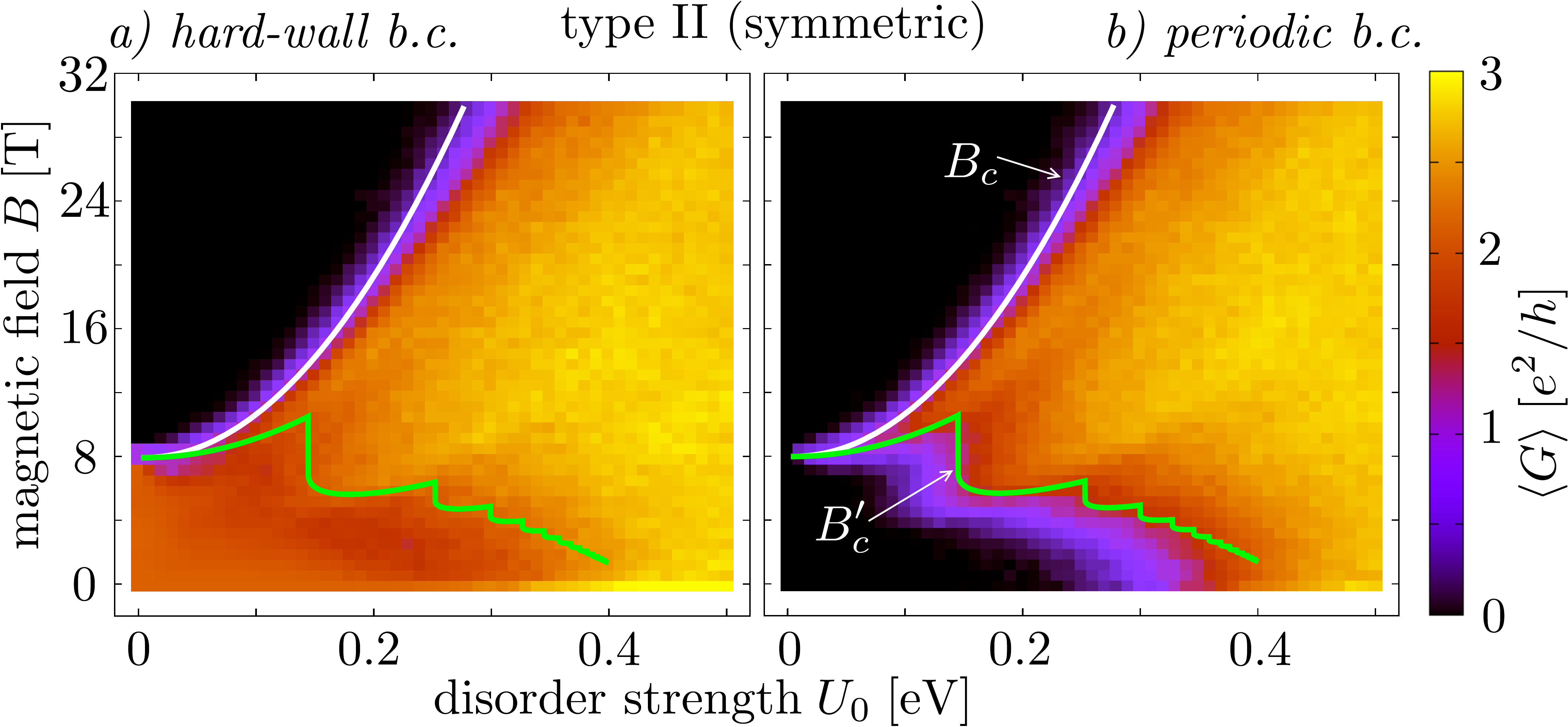}}
\caption{Disorder-averaged conductance of a type II quantum well with electron-hole symmetry ($\delta\mu=0$, other parameters as in Table \ref{Tablequantumwell}), calculated numerically from the tight-binding Hamiltonian \eqref{calHdef}, with hard-wall boundary conditions (panel \textit{a}) or periodic boundary conditions (panel \textit{b}). The Fermi level is set at $E_{\rm F}=8\cdot 10^{-4}\,{\rm eV}$, slightly displaced from the center of the bulk gap to avoid the minigap in the spectrum of helical edge states. The disorder dependence of the characteristic fields $B_c$ and $B'_c$ (white and green curves) is calculated from the renormalization of the band gap in Born approximation, as described in the text.}
\label{fig_typeIIphase}
\end{figure*}

\begin{figure*}[tb]
\centerline{\includegraphics[width=0.8\linewidth]{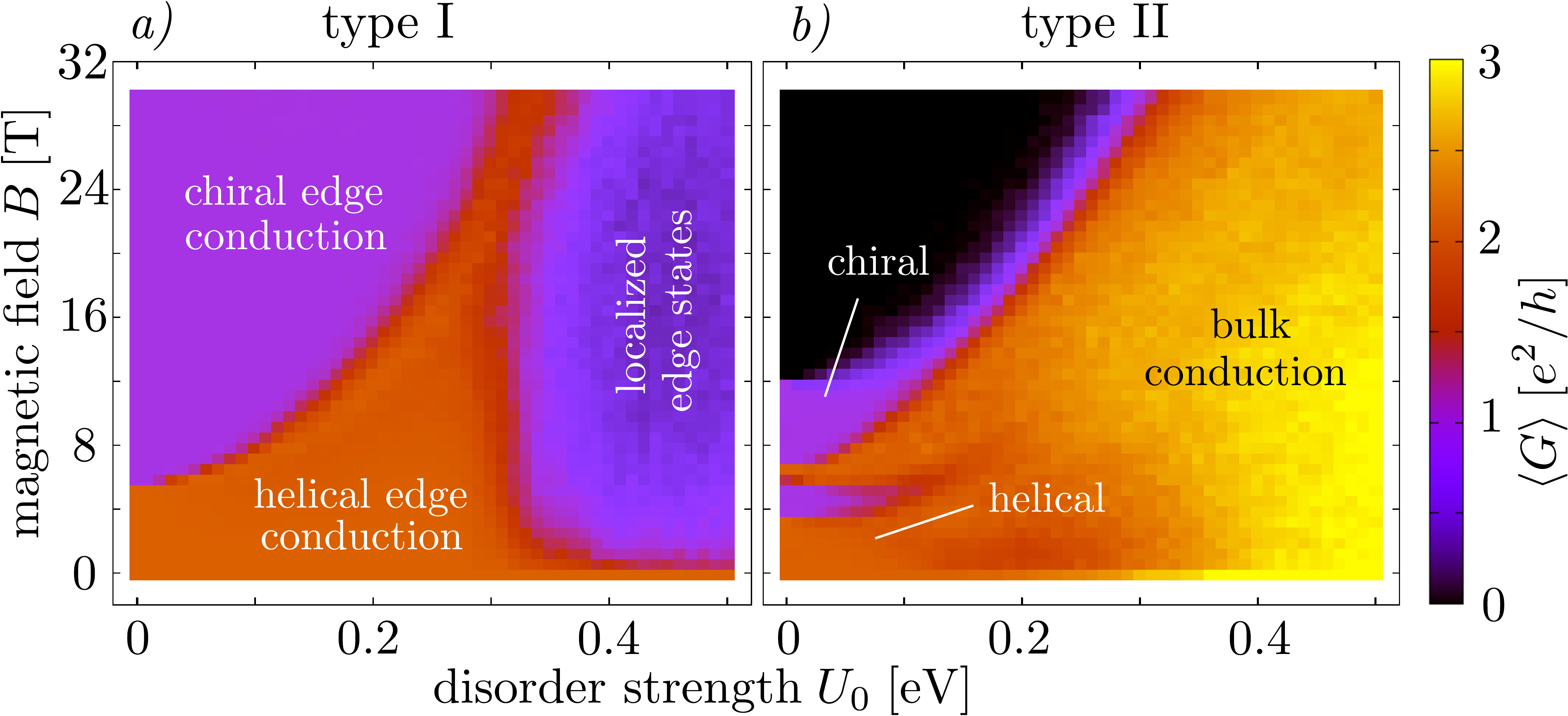}}
\caption{Disorder-averaged conductance of a quantum well of type I (panel \textit{a}) and type II (panel \textit{b}), with hard-wall boundary conditions. The parameters are those of Table \ref{Tablequantumwell}, including the effects of broken electron-hole symmetry (nonzero $\delta\mu$). The conductance is calculated at the renormalized Fermi energy \eqref{EFresult}. The region of bulk conduction is present in the type II quantum well, but not in type I, where instead a region of localized edge states appears. (This region turns black for periodic boundary conditions, so we know there is no bulk conduction there.)}
\label{fig_typeItypeIIphase}
\end{figure*}

To study the effect of disorder we discretize the tight-binding Hamiltonian \eqref{calHdef} on a square lattice (lattice constant $a=2.5\,{\rm nm}$, size $W\times L =  500\,{\rm nm}\times300\,{\rm nm}$). Randomly distributed dopants are introduced by adding a spin- and layer-independent random potential $U(\bm{r})$, fluctuating from site to site in the interval $(-U_0/2,U_0/2)$. We take either periodic or hard-wall boundary conditions along the sides at $y=0,W$ and attach the ends at $x=0,L$ to ballistic leads to obtain the transmission matrix $t$ at the Fermi level $E_{\rm F}$. The conductance $G=(e^2/h)\,{\rm Tr}\,tt^{\dagger}$ is averaged over 60 disorder realizations. All calculations were performed using the {\sc kwant} tight-binding code.\cite{kwant} Results are presented in Figs.\ \ref{fig_typeIIphase} and \ref{fig_typeItypeIIphase}.

We first discuss Fig.\ \ref{fig_typeIIphase}, which shows data for the type II quantum well with electron-hole symmetry. The QSH regime of helical edge conduction appears as a region of quantized conductance $G=2e^{2}/h$ in the low-field/weak-disorder corner of panel \textit{a} (hard-wall boundary conditions). The high-field/weak-disorder corner is the QH regime, with $G=0$ because the Fermi level lies in the gap between the chiral edge states of conduction and valence band. The region between the QSH and QH regimes has a nonquantized conductance $G\gtrsim 2e^{2}/h$. This is a regime of bulk conduction, since a removal of the edge states by switching from hard-wall to periodic boundary conditions makes no difference (compare panels \textit{a} and \textit{b}). We call this regime ``quasi-metallic'' rather than metallic, because in the limit of an infinite system all bulk states should localize in a magnetic field.

The curves marked $B_c$ and $B'_c$ in Fig.\ \ref{fig_typeIIphase} are obtained as in Fig.\ \ref{fig_LL2}, with the effects of disorder accounted for as follows: We replace the zero-field band gap $M_0$ by the renormalized gap $M_{\rm eff}(U_0)$ in Born approximation,
\begin{equation}
M_{\rm eff}=M_{0}-c\,U_0^2,\label{Meffresult}
\end{equation}
with $c=0.39\,[{\rm eV}]^{-1}$. The band gap $M_0$ of the clean system is negative, so disorder increases the band gap, as in the topological Anderson insulator.\cite{Gro09} There is no renormalization of the Fermi energy for $\delta\mu=0$. The Landau level broadening is estimated at $\delta E=U_0^2\times 0.05\,[{\rm eV}]^{-1}$, so that the characteristic fields are determined by the first and last Landau level crossing with the line $E=\delta E$ (rather than with $E=0$). As is evident from Fig.\ \ref{fig_typeIIphase}, the resulting curves $B_c(U_0)$ and $B'_c(U_0)$ describe quite well the boundaries of the quasi-metallic regime, over a broad range of magnetic fields and disorder strengths.

These are results for the electron-hole symmetric case $\delta\mu=0$, but the appearance of the magnetic-field induced quasi-metallic regime is a generic feature of inverted type II quantum wells, not tied to electron-hole symmetry --- the weak electron-hole coupling is the essential ingredient. This is demonstrated in Fig.\ \ref{fig_typeItypeIIphase}. Because of the non-zero $\delta\mu$ the Fermi energy is renormalized by disorder, which we take into account in Born approximation,
\begin{equation}
E_{\rm{F}}=-d\,U_0^2.\label{EFresult}
\end{equation}
The coefficient $d$ equals $0.12\,[{\rm eV}]^{-1}$ and $0.27\,[{\rm eV}]^{-1}$, respectively, in the type I and type II quantum wells. 

Comparing the results for the type II quantum well (with hard-wall boundary conditions), we see that Figs.\ \ref{fig_typeIIphase}\textit{a} and \ref{fig_typeItypeIIphase}\textit{b} are qualitatively similar, the main effect of the broken electron-hole symmetry being the appearance at weak disorder of a regime of quantized chiral edge conductance ($G=e^2/h$). As one can see in Fig.\ \ref{fig_LL1}\textit{b}, the Landau levels depend nonmonotonically on the magnetic field, and this shows up in Fig.\ \ref{fig_typeItypeIIphase}\textit{b} as a nonmonotic variation of the conductance from $2\rightarrow 1\rightarrow 2\rightarrow 1\rightarrow 0\times e^2/h$ at weak disorder. 

Both figures \ref{fig_typeIIphase}\textit{a} and \ref{fig_typeItypeIIphase}\textit{b} show the regime of bulk conduction at strong disorder characteristic of an inverted type II quantum well. This regime requires small electron-hole coupling, to allow for an accumulation of Landau levels near the Fermi energy (compare Figs.\ \ref{fig_LL1}\textit{a} and \ref{fig_LL1}\textit{b}). For that reason the quasi-metallic regime is absent in the type I quantum well of Fig.\ \ref{fig_typeItypeIIphase}\textit{a}, which instead shows the expected\cite{Mac10,Del12} transition to localized edge states at strong disorder.

\begin{figure}[tb]
\centerline{\includegraphics[width=0.9\linewidth]{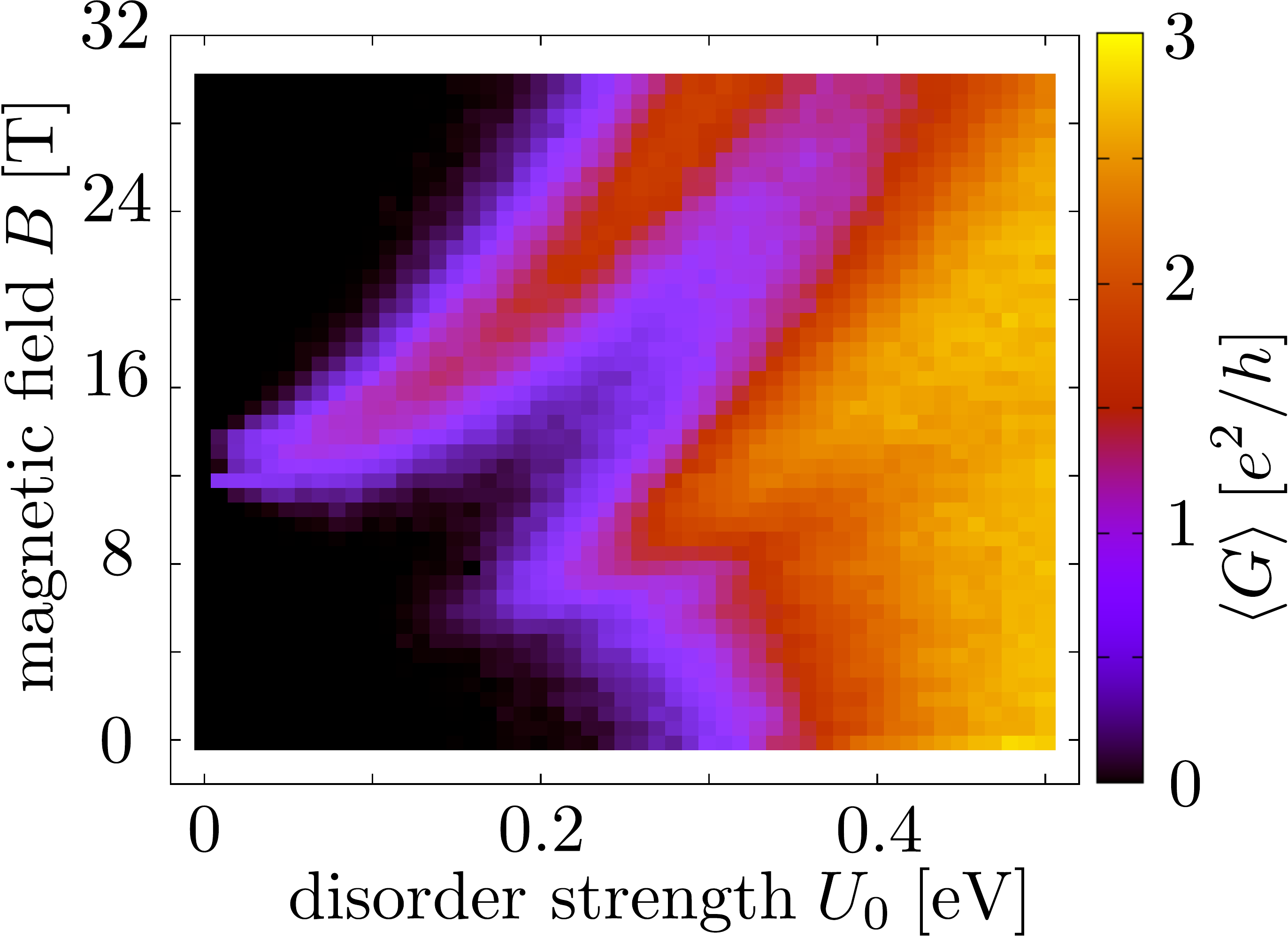}}
\caption{Same as Fig.\ \ref{fig_typeIIphase}\textit{b}, but including the effects of the Zeeman energy with effective $g$-factor $-7.5$.}
\label{fig_typeIIg15}
\end{figure}

So far we have focused on the orbital effect of a perpendicular magnetic field. The effect of spin splitting by the Zeeman energy $g\mu_{\rm B}B$ is shown in Fig.\ \ref{fig_typeIIg15}, for the type-II quantum well with periodic boundary conditions and electron-hole symmetry. We took the value $g=-7.5$ of bulk InAs, comparable absolute values may be expected for a narrow InAs/GaSb quantum well.\cite{Smi87} Comparison with Fig.\ \ref{fig_typeIIphase}\textit{b} (where we had $g=0$) shows a qualitatively similar phase diagram, in particular the regime of bulk conduction persists. Possible due to electron-electron interactions or geometry larger g-factors may have more dramatic effect.\cite{Pik14}

In summary, we have investigated how disorder affects the breakdown of the QSH effect in a perpendicular magnetic field. In inverted type I quantum wells, such as HgTe, the characteristic breakdown field $B_c$ increases with disorder strength, due to a renormalization of the band gap (becoming more and more negative with increasing disorder). The same effect is operative in broken gap quantum wells of type II, such as InAs/GaSb --- however, there it does not lead to an increased robustness of helical edge conduction. The spatial separation of the inverted electron-hole subbands leads to the accumulation of Landau levels in the zero-field band gap, producing a regime of bulk conduction that extends to lower and lower magnetic fields with stronger disorder, see Fig.\ \ref{fig_typeIIphase}.

One implication of our findings, see Fig.\ \ref{fig_typeItypeIIphase}, is that the weak disorder limit is in principle consistent with the persistence of helical edge conduction up to 8\,{\rm T} perpendicular fields, reported in Ref.\ \onlinecite{Du13}. However, in the presence of  strong disorder the bulk conduction is expected to take over at much smaller fields.

As directions for further research, it would be interesting to explore the fate of the quasi-metallic regime in the thermodynamic limit. All two-dimensional bulk states should localize in a magnetic field, but the numerics suggests a large localization length. It would also be of interest to study the effect of Landau level accumulation on exciton condensation in the electron-hole bilayer, considered recently in connection with the QSH effect.\cite{Pik13,Bud13}  

We are grateful to Jason Alicea for useful discussions. This research was supported by the Foundation for Fundamental Research on Matter (FOM), the Netherlands Organization for Scientific Research (NWO/OCW), an ERC Synergy Grant, and the China Scholarship Council.


\begin{thebibliography}{99}
\bibitem{note1} Following Ref.\ \onlinecite{Liu08b}, the type I parameters in Table \ref{Tablequantumwell} are those of a $7\,{\rm nm}$ thick HgTe quantum well. The type II parameters correspond to a $8.4\,{\rm nm}/10\,{\rm nm}$ InAs/GaSb quantum well, except for $M_0$, for which we took the same value as for the type I quantum well in order to simplify the comparison.
\end{thebibliography}
\end{document}